# Introducing recalibrated academic performance indicators in the evaluation of individuals' research performance: A case study from Eastern Europe

**György Csomós**
University of Debrecen
2-4 Otemeto u., Debrecen 4028, Hungary

**Abstract**

In Hungary, the highest and most prestigious scientific qualification is considered to be the Doctor of Science (DSc) title being awarded by the Hungarian Academy of Sciences. The academic performance indicators of the DSc title are of high importance in the evaluation of individuals' research performance not only when a researcher applies for obtaining a DSc title, but also during promotions and appointments at universities, and in the case of the evaluation of applications for scientific titles and degrees, and the assessment of applications for funding. In the Section of Earth Sciences encompassing nine related disciplines, rather than carrying out a straightforward bibliometric analysis, the performance indicators were designed as a result of a consensual agreement between leading academicians, each of whom represented a particular discipline. Therefore, the minimum values of the indicators, required to be fulfilled if one is applying for a DSc title, do not adequately reflect the actual discipline-specific performance of researchers. This problem may generate tension between researchers during the evaluation process. The main goal of this paper is to recalibrate the minimum values of four major performance indicators by taking the actual discipline-specific distance ratios into account. In addition, each minimum value will be defined by employing integer and fractional counting methods as well. The research outcome of this study can provide impetus for the Section of Earth Sciences (and eventually other sections of the Hungarian Academy of Sciences) to optimize the minimum values of the DSc title performance indicators by taking the specifics of each discipline into account. Because academic performance indicators are also employed in other Eastern European countries in the evaluation of individuals' research performance, the methods used in that paper can be placed into a wider geographical context.

## 1. Introduction

The political, social and economic changes that emerged in the early 1990s enabled the Central and Eastern European (CEE) countries to become more integrated into the European and global economy. The transition phase is seen as an important turning point in the development of science systems in the CEE countries because science has freed itself from the indirect political and ideological control of the Soviet Union (Kozak, Bornmann & Leydesdorff, 2015). During the transition, most CEE countries introduced reforms in their higher education system by adopting the Bologna process (Kozma, 2014), and also in their academic qualification system by introducing a PhD degree, reflecting the qualification scheme applied in Western Europe (Taylor, Kiley & Humphrey, 1998). Prior to the adoption of the PhD degree, varying types of scientific qualifications were used in the region, most of them imported from the Soviet Union. Universities were allowed to award "university doctor" (dr. univ.) and Habilitation (habil.) titles, whereas scientific academies were authorized to award higher qualifications such as "candidate of science" (C.Sc.) and "doctor of science" (D.Sc.) degrees (Hangos, 1997; Quandt, 2002). In the new system, the universities were provided with the right to award the PhD degree, and the previous rigid hierarchy of scientific qualifications was broken down. Some CEE countries, however, are unique in that they are characterized by a sort of "dualistic" scientific qualification scheme because, beside the internationally acknowledged PhD degree, they still allocate more or less significance to some of the old qualifications as well. In the Czech Republic, the DSc degree, a higher qualification than the PhD, is awarded by the Czech Academy of Sciences, and in Poland, the Habilitation is the highest academic qualification (Korytkowski & Kulczycki, E., 2019). Whereas PhD degrees are awarded on the basis of a thesis reviewed by independent researchers, the evaluation of Habilitation and DSc degrees is more metric based (Kulczycki, 2017; Kulczycki, Korzeń & Korytkowski, 2017). However, both in Poland and Hungary, two of the largest scientific actors in the CEE region (Pajić, 2015), there are constant debates regarding the use of bibliometric indicators in the evaluation of an individual's research performance.

In Hungary, the highest and most prestigious scientific qualification is considered to be the Doctor of Science (DSc) title. The DSc is awarded by the Hungarian Academy of Sciences (HAS), whereas the PhD degree and the Habilitation (which is also a title) are awarded by universities. To obtain a DSc title, university teachers and researchers are required to fulfill much higher performance indicators than is necessary to acquire a PhD degree. Until recently, most universities had not allowed associate professors and college professors to apply for a university professor position (which is considered to be the top of the academic career ladder) until they had obtained a DSc title. Today, parallel with the weakening position of the HAS in the Hungarian science system, the DSc title has lost its significance as being the fundamental criterion of the university professorship. Moreover, the DSc title has been replaced with the Habilitation title, a qualification being awarded by universities, and is now considered to be the highest qualification required to obtain a university professor position. The Hungarian Accreditation Committee (HAC), a national-level, independent body of experts tasked with the external evaluation of applications for awarding university professor positions, considers the DSc title to be only an advantage rather than a fundamental criterion. That is, it appears to be no longer important whether university teachers and researchers have a DSc title if applying for university professor positions because the promotion process has become the exclusive competence of the HAC and the universities. However, through the definition of the academic performance indicators being components of the broader requirements of the DSc title, the HAS significantly influences not only the academic career advancement of individuals but also the main features of the national-level scientific evaluation processes. The reason for this is that both in the case of promotions and appointments at universities, and the evaluation of an individual's application for scientific qualifications (i.e., the PhD degree and the Habilitation title), a specific proportion of the performance indicators' minimum values must be taken into account. In addition, the performance indicators of the DSc title are incorporated into the scientific requirements of national funding programs (e.g., the "OTKA", the most important basic research program, coordinated by the National Research, Development and Innovation Office) and scholarships (e.g., the János Bolyai research scholarship available for young researchers provided by the HAS).

In conclusion, for university teachers and researchers, it is now not considered a fundamental requirement to obtain a DSc title if applying for a higher position; however, in one way or another, the DSc title performance indicators issued by the HAS will definitely impact their career path.

The HAS did not introduce standardized performance indicators but allowed its 11 scientific Sections, each of which represents a broader scientific field (e.g., agricultural sciences, engineering, and medical sciences), to develop customized indicators. For this reason, the overall composition of the performance indicators and the minimum value of those indicators vary Section to Section. Moreover, due to the fact that each Section encompasses a wide range of scientific disciplines having different publication characteristics and output, the types and values of the performance indicators can vary even within a particular Section. For example, the Section of Engineering Sciences hosts 15 engineering-related disciplines (each of which is represented by a scientific committee) out of which the publication characteristics of the discipline "architecture" is quite different from that of the discipline "material sciences" (i.e., the differences of the performance indicators should reflect the differences of the publication characteristics of disciplines). However, irrespective of how significant the differences are between the publication characteristics of particular disciplines or groups of disciplines, the Section does not allow (more precisely: cannot allow) its committees to employ such performance indicators that perfectly reflect those differences. If committee "A" wants to use overly different indicators from the ones used by other committees, it might suggest that committee "A" is most probably not in the right Section. Thus, the types and values of the performance indicators must reflect the unitedness of the Section. In some Sections, however, this sort of forced closeness of the performance indicators may generate tension between researchers because this procedure ignores the actual differences of the publication characteristics and output of researchers affiliated with particular disciplines (i.e., in some extreme cases, there is an attempt to compare apples with oranges).

Naturally, the selection of the most adequate performance indicators seems to be a significant challenge worldwide (Coomes, Moore, Paterson, Breau, Ross, & Roulet, 2013; Sahel, 2011; Schreiber, Malesios, & Psarakis, 2012; Wildgaard, Schneider, & Larsen, 2014). The most critical factors are deemed to be the following ones: whether the quality or quantity should be considered to be more important (Bucur, Kifor, & Mărginean, 2018, Kallio, Kallio, & Grossi, 2017), how much importance should be given to the impact factor (Brito & Rodríguez-Navarro, 2019; McKiernan, Schimanski, Nieves, Matthias, Niles, & Alperin, 2019; Zhang, Rousseau, & Sivertsen, 2017), which of the indexing databases (e.g., Web of Science, Scopus and Google Scholar) should be used as the basic source during the evaluation process (Martín-Martín, Orduna-Malea, Thelwall, & Delgado López-Cózar, 2018; Mikki, 2010; Vieira & Gomes, 2019), and which of the counting methods (e.g., integer, fractional, and first author counting) should be employed (Egghe, Rousseau, & Van Hooydonk, 2000; Gauffriau, Larsen, Maye, Roulin-Perriard, A., & Von Ins, 2007; Gauffriau, 2017; Van Hooydonk, 1997).

The above questions, of course, are also raised in Hungary during the evaluation of individuals' research performance. In addition, due to some external factors (e.g., the increasing popularity of open-access publishing, the temporal inaccessibility of the Scopus and other Elsevier products, and the decreasing funding of basic research), the reform of the academic performance indicators seems unavoidable.

In this study, a bibliometric analysis is conducted to examine the quality and quantity of the publication output of individuals affiliated with the Hungarian Academy of Sciences, Section of Earth Sciences in the period of 2014−2018. By combining hard natural science disciplines (e.g., geophysics, geochemistry, and meteorology) and a social science discipline (i.e., social geography[1]) (see, Coomes et al., 2013) under one roof, the Section of Earth Sciences is considered to be one of the most special sections of the HAS. In the Section, the discipline-specific publication characteristics significantly differ, but the differences are not correctly taken into account. As a consequence, the research performance of individuals is evaluated incorrectly and in a less fair manner. The most significant problem stems from the fact that the current minimum values of the performance indicators (as defined by the scientific committees) are not in line with the real performance values (as can be experienced in reality). According to a piece of informally obtained information, the minimum values were created as the outcome of a consensus agreement of a couple of senior researchers (each of whom had had a DSc title), rather than carrying out a straightforward bibliometric analysis.

---

[1] According to the generally employed international classification of scientific disciplines, social geography is considered to be a branch of human geography, a scientific field that is most commonly referred to simply as geography. In addition, the discipline of geography (which is distinguished from physical geography) is classified as a social science branch; therefore, except in Hungary, it is not considered to be a subfield of Earth Sciences.

This study, by systematically analyzing the publication output of researchers affiliated with the Section of Earth Sciences, proposes an alternative method regarding how to most optimally recalibrate the DSc title's performance indicator minimum values. In addition, the method employed in this paper can serve as an example for the HAS itself, and eventually other CEE countries whose research evaluation system is similarly highly metric based.

**2. Data and methods**

**2.1. The Section of Earth Sciences in brief**

The Section of Earth Sciences encompasses the following disciplines (listed in alphabetical order): geochemistry, geodesy, geology, geophysics, meteorology, mineralogy, mining, paleontology, petrology, physical geography, and social geography (https://mta.hu/english/scientific-sections-105963). The Section consists of 11 scientific committees out of which two (the Committees of Anthropology and Microbiology) are so-called intersectional scientific committees belonging to two sections at the same time. The disciplines being represented by the intersectional scientific committees are out of the scope of this analysis because the performance indicators of the DSc title do not pertain to those disciplines. That is, the analysis involves nine disciplines, each represented by a particular committee:
- Committee on Geochemistry, Mineralogy, and Petrology
- Committee on Geodesy
- Committee on Geography I (Social Geography)
- Committee on Geography II (Physical Geography)
- Committee on Geology
- Committee on Geophysics
- Committee on Meteorology
- Committee on Mining
- Committee on Paleontology

According to the public database of the HAS containing personal information on researchers, as of August 31, 2019, a total number of 805 individuals were affiliated with the Section of Earth Sciences.

It is necessary to note that in the study, the terms "committee" and "discipline" are used as quasi synonyms. That is, it is supposed, for example, that a given researcher being affiliated with the Committee on Paleontology, conducts research in the field of paleontology and produces publications in that field. Clearly, in reality, the above logical relationship is not necessarily true (e.g., paleontologists publish papers that can be classified to the field of geology as well); however, in most cases, the authors do not indicate the discipline into which their publication should be classified.

**2.2. The Hungarian Scientific Bibliography, the national publication and citation database**

In this study, the Hungarian Scientific Bibliography (HSB) is used to map bibliometric data. The HSB, launched in 2009, is a comprehensive bibliographic database of scientific publications produced by Hungarian researchers, and the citations those publications receive (Holl, Makara, Micsik, & Kovács, 2014). One major advantage of the HSB is that it stores the bibliographic data of any types of publication written by Hungarian researchers in any language. In some fields, particularly in arts and humanities, but also in social sciences, most of the publications are written in Hungarian and are not indexed in international databases. That is, the HSB helps researchers find each other's publications in an organized manner. In addition, for funders and policy-makers, the HSB provides improved transparency on how effective the use of grants and public money are. It must be noted, however, that the HSB provides less optimal conditions for conducting bibliometric analysis than such prestigious abstract and citation databases as Web of Science (WoS) and Scopus. Thus, it is necessary to present some limitations regarding the HSB.

Naturally, in international contexts, WoS and Scopus are the most generally used databases to conduct bibliometric analysis in the case of earth sciences (see, for example, Coomes et al., 2013; Gorraiz, Gumpenberger, & Glade, 2016, Rey-Rocha & Martín-Sempere, 2004; Wang & Liu, 2014). One of the most important advantages of WoS, owned by Clarivate Analytics, is that it provides such customized, citation-based research analytics tools as the InCites and Essential Science Indicators platforms. The utilization of WoS for conducting this analysis would be a quite reasonable choice if we did not realize two fundamental problems (at least viewed from the perspective of Hungarian researchers): 1) nearly two-thirds (64 percent) of the publications indexed in WoS are journal articles, and 2) English-language publications are significantly overrepresented in the database (Mongeon & Paul-Hus, 2016). The latter factor seems to be rather problematic because, due to the heavy language bias of WoS, a more favorable condition is created for natural sciences against social sciences in which a significant proportion of the publications are produced in a non-English language. For example, in the period of 2014−2018, only 20 percent of the publications produced by Hungarian researchers affiliated with the Section of Earth Sciences were indexed in WoS (this ratio is as low as 7.2 percent in the case of social geography).

Scopus, an abstract and citation database of Elsevier, offers similar bibliometric analytical tools to those provided by WoS, but it covers a much larger publication portfolio. The content of Scopus is also dominated by the English language, but not to such a degree as can be experienced in the case of WoS. This feature of Scopus makes it more suitable for conducting bibliometric analysis in the fields of social sciences and humanities (Archambault, Vignola-Gagné, Côté, Larivière, & Gingrasb, 2006). However, from the perspective of this analysis focusing on evaluating the publication performance of Hungarian researchers, Scopus has a similar problem to WoS: in its dataset, English-language journal articles are overrepresented. In addition, at the end of 2018, due to the fact that Elsevier did not address the requirements of the Hungarian Negotiation Committee, the negotiation process between Hungary and Elsevier was terminated. Hence, at the beginning of 2019, Scopus, and all other Elsevier products, were temporarily unavailable for universities. Finally, a deal has been made; however, it is not guaranteed that the contract will be renewed (a long-term subscription to WoS seems to be more realistic).

In conclusion, irrespective of how user friendly WoS and Scopus are, considering the special circumstances, there is no other option but to employ the HSB. The HSB contains data on various types of publications written in any languages (Holl et al. 2014) and provides summary statistics on researchers' publication performance. Therefore, the HSB seems to be the optimal (more precisely: the only possible) choice to conduct bibliometric analysis regarding the publication performance of Hungarian researchers. In addition, the HSB has a highly critical feature: the bibliographic data of publications must be uploaded voluntarily. In contrast to WoS and Scopus, in the case of the HSB, the data upload and processing are not handled by a professional team but are the duty and responsibility of the researchers (i.e., the authors) themselves. This procedure generates (at least) two major problems: First, many researchers (primarily the senior researchers) simply do not have an HSB profile, and even if they produce publications (which is quite likely), they do not have a trace in the HSB. Second (and representing a more problematic factor than the previous one), the credibility of bibliographic data of publications and citations uploaded voluntarily by researchers into the HSB is questionable, to say the least. Before initiating an evaluation of an application for the DSc title, the HAS temporarily blocks the use of the applicant's HSB profile to carefully review the correctness of his/her publications' bibliographic data (in fact, according to a piece of information provided by a leading librarian, only a randomly chosen 10 percent of the publications are required to be reviewed). Until a "thorough" revision initiated by the HAS, the correctness of bibliographic data of publications uploaded voluntarily by authors into the HSB is checked by a local administrator (i.e., most typically a librarian who is affiliated with the host institution), who does not have sufficient time to carefully review each publication and citation data one by one (in some cases, they lack professional experience as well). That is, utilization of the HSB for conducting this bibliometric analysis is only motivated by the fact that a considerably high ratio of publications produced by Hungarian researchers (particularly that of social geographers) is not indexed in either WoS or Scopus. During the analysis, the limitations of the HSB should be kept in mind.

The analysis involves publication data from the period of 2014−2018. Naturally, when a researcher applies for a DSc title, the bibliometric analysis to reveal his/her individual publication performance focuses on his/her full publication history and not only the data extracted from a short time period. However, the main goal of this study is to recalibrate the DSc title's performance indicator minimum values; therefore, the differences between the publication performance of young and senior researchers must be balanced by observing them in a reasonable time period (i.e., it is highly likely that senior researchers have produced more publications; consequently, their bibliometric data will be much higher). In contrast to the publication data, the time period of citation data ranges from 2014 to 2019 because the HSB does not make it possible to adjust the time interval of the citation search (the data collection was completed on December 31, 2019).

In the period of 2014−2018, out of the total number of 805 members affiliated with the Section of Earth Sciences, 569 researchers had a profile and publication record in the HSB. These researchers produced 11,960 publications during that period, and those publications received 20,702 independent citations (see more thorough explanations on independent citations in Sections 2.4 and 3.3).

**2.3. DSc title's academic performance indicators**

Similar to other Sections of the HAS, the Section of Earth Sciences maintains a Doctoral Committee, the procedure of which contains the minimum values of the performance indicators required to be achieved if an individual intends to apply for a DSc title. For applicants, it is essential to fulfill the minimum value of each performance indicator (Table 1), but it is highly recommended that they be over-fulfilled (e.g., a researcher who fulfills the minimum value of an indicator receives one point, but if he/she proportionally over-fulfill the given indicator, he/she receives proportionally higher points). Each discipline with similar publication characteristics (at least it is supposed that they have similar publication characteristics) have been merged into a single group; that is, the nine disciplines of the Section have been classified into three groups.

**Table 1.** Minimum performance indicator values by groups of disciplines as defined by the Section of Earth Sciences

|  | Disciplines in Group 1: geochemistry, mineralogy, petrology, geology, geophysics, meteorology, and paleontology | Disciplines in Group 2: mining, geodesy, geoinformatics, and physical geography | Discipline in Group 3: social geography |
|---|---|---|---|
| Number of scientific publications | 30 | 30 | 40 |
| Number of scientific publications with first author position | 15 | 15 | 20 |
| Number of scientific publications since obtaining last scientific degree | 15 | 15 | 30 |
| Number of scientific books and monographies | - | - | 2 |
| Number of scientific publications published in a foreign language | - | - | 35 |
| Number of journal articles indexed in SCI/SSCI and Scopus | 12 | 8 | 6 |
| Number of journal articles indexed in SCI/SSCI and Scopus since obtaining last scientific degree | 6 | 4 | 3 |
| Number of independent citations | 150 | 120 | 150 |
| Number of independent citations located in SCI/SSCI and Scopus | 50 | 30 | - |
| Cumulative impact factor value | 8 | 4 | 2 |
| Hirsch index | 9 | 8 | 8 |

As can be observed in Table 1, the social geography discipline forms a single group on its own, and requests two additional indicators from the applicants to be revealed: the number of scientific books and monographies, and the number of scientific publications published in a foreign language. Both indicators refer to the social science orientation of the discipline (the significance of books as publication type in social sciences is examined by, for example, Hicks, 1999; Larivière, Archambault, Gingras, & Vignola-Gagné, 2006), and this fact is also underpinned by the low number of SCI/SSCI and Scopus articles and the low cumulative impact factor values being requested from the applicants.

    Furthermore, it should be noted, that in the case of social geography, the Section requests from the applicants to present the total number of articles published in SSCI- and Scopus-indexed journals. In the case of other disciplines, the researchers have to demonstrate the total number of journal articles indexed in both SCI and Scopus. This phenomenon should be highlighted due to two reasons: First, it is unclear why researchers engaged in social geography are not supposed to indicate those articles that have been published in SCI journals (from an opposite perspective, this problematic situation is true regarding the remaining researchers affiliated with other disciplines). Naturally, it is also possible to demonstrate the number of articles indexed in Scopus, making the above problem almost irrelevant. According to Gavel and Iselid (2008), there is a significant overlap between the contents of Scopus and WoS; for example, in 2016, 84 percent of active titles in WoS were also indexed in Scopus. Second, it is true that the HSB provides section-specific summary statistics per author, but the problem is that there is a difference between the indicators the HSB demonstrates and those that the Section requests. In fact, rather than demonstrating the number of SCI and SSCI journal articles authored by the researchers, the HSB presents information on the number of articles being extracted from the entire WoS database (the data type of which, therefore, does not meet with the ones requested by the Section). In 2015, the WoS launched the Emerging Sources Citation Index (ESCI), a new index to include peer-reviewed publications of regional importance and in emerging scientific fields. Due to this development, the "Tér és Társadalom" (Space and Society), a journal publishing articles in Hungarian with abstracts in English, has been selected to be included in the ESCI, and thus in the WoS Core Collection. The "Tér és Társadalom", a popular journal for Hungarian social geographers, is not indexed in the Scopus database nor in SCI/SSCI. Due to the fact, however, that the articles published in the "Tér és Társadalom" are included in the WoS Core Collection (and are equipped with a WoS Accession Number), they increase the value of the performance indicator "Number of journal articles indexed in SCI/SSCI and Scopus", irrespective of the fact that in reality, those articles are not indexed in SCI/SSCI nor in Scopus. That is, when a researcher affiliated with the Section of Earth Sciences submits an application for obtaining a DSc title (or the university professor position), he/she considers the bibliometric data of his/her full WoS record as demonstrated by the HSB, which in fact does not meet the requirements of the Section.

This study employs bibliometric data provided by the HSB; that is, when conducting the bibliometric analysis, the number of WoS articles and independent citations are considered. In addition, due to three reasons, a constraint must be implemented regarding the utilization of the Scopus database: 1) Due to the fact that the ESCI (containing approximately 7,800 titles) has been launched, the overlap between the contents of both WoS and Scopus has become much higher. 2) If our aim is to obtain information on researchers' WoS and Scopus records in a particular time period, the HSB does not allow us to collect data automatically. Therefore, each record (approximately 12,000 publications and 20,000 independent citations) should be scrutinized manually (which is, in fact, rather time consuming). However, if we consider only one of the identifiers, by employing a semi-automatic search method, the search process can be accelerated. 3) In contrast to WoS, the long-term subscription to Scopus seems uncertain.

**2.4. Methods**

To conduct the analysis, the publications authored by 569 researchers affiliated with the Section of Earth Sciences in the period of 2014−2018 were scrutinized one by one. In addition, when collecting information on independent citations (i.e., citations excluding self-citations and citations received from co-authors) that those articles received, the period of 2014−2019 was taken into account. As of December 31, 2019, based on data provided by the HSB, research affiliated with the Section produced a total number of 11,960 publications, and those publications received 20,702 independent citations. From among the 11 performance indicators, four have been chosen to be examined. The reason for this is that each of the four indicators can be found in the requirements list of the three discipline-groups, respectively; moreover, the HSB provides adequate and comparable information on those indicators. These indicators are as follows: "The number of scientific publications", "The number of journal articles indexed in the WoS", "The number of independent citations", and the "The cumulative impact factor value".

For each researcher, the value of a particular indicator was detected. Based on the individual publication performance per indicator, a dataset was compiled. To recalibrate the value of a given performance indicator, only the top 25 percent of individual publication performance was considered. For example, the number of researchers affiliated with the Committee on Paleontology and with an HSB profile was 32; that is, in the case of each indicator, only the highest eight individual performances (i.e., the top 25 percent of individual publication performance) were involved in the analysis. Currently, as it is suggested by a piece of informally obtained information, the minimum values of the performance indicators reflect on the performance of scholars having a DSc title. However, after scrutinizing the dataset, it turned out that if we focus on a particular time period (in this case, 2014 to 2018), the performance of many researchers with a DSc title rather approximates the average. Therefore, to introduce a higher standard for future candidates, a new approach is necessary.

In addition, in the case of each indicator, the discipline-specific minimum values were defined by using both integer and fractional counting methods. When employing fractional counting, one credit is equally shared among the co-authors of a given publication; then, the fractional credit values are summarized per author.

More precisely, the method is as follows: Regarding each discipline and indicator, the average of the top 25 percent of individual publication performance was calculated. In such a manner, we could obtain information on the discipline-specific actual performance values (APVs). By determining the APVs, in the case of each discipline, it became possible to demonstrate the number of years required to achieve the discipline-specific current minimum values (CMVs) in reality.

$$Y_i = \frac{CMV_i}{APV_i} * t \tag{1}$$

where,

$Y_i$ = the number of years necessary to fulfill the CMV of a given indicator if taking the APV into account
$CMV_i$ = the discipline-specific current minimum value regarding a given indicator
$APV_i$ = the discipline-specific actual performance value regarding a given indicator
$i$ = a particular discipline
$t$ = the time horizon of the data employed (in this case, it is 5 years)

Naturally, the number of years required to fulfill the CMV of a given indicator varies from discipline to discipline. For example, in the discipline of geology, the average of the top 25 percent of individual publication performance value is 48.769 publications during 5 years, whereas the top researchers in the discipline of geodesy/geoinformatics produce an average number of 30.900 publications in the same period. In fact, the CMV for both disciplines is 30 publications, respectively. That is, a top geologist (defined on the basis of the average of the top 25 percent of individual publication performance) can fulfill the CMV regarding this particular indicator (i.e., the number of scientific publications) during 3.076 years (by producing an average number of 9.754 publications per year), whereas for a top researcher in geodesy, it takes 4.854 years (6.180 publications per year). Let us approach this issue from a different perspective: Taking a 15-year period into account, a top geologist produces 146 publications, in contrast to a top surveyor, who writes 93 publications. That is, when a geologist and a surveyor of the same age and same position in terms of individual publication performance (i.e., both of them belong to the top 25 researchers in the discipline they are affiliated with) applies for funding (e.g., an OTKA basic research grant), they might have quite different chances to win because their publication output significantly differs (yet, both researchers will be evaluated according to the same standards).

To create a balance between the disciplines, the minimum values of the performance indicators must be recalibrated by taking the actual publication characteristics of each discipline located in the Section into account. It is also important that the discipline-specific distance ratio (DSDR), that is, the ratio that reflects the differences between the APVs regarding a particular discipline must be incorporated into the recalibrated minimum values. Moreover, the CMV of the performance indicators must also be taken into account.

In conclusion, to introduce the recalibrated minimum values (RMVs), both current and actual DSDRs must be calculated.

$$DSDR_{c,i} = \frac{CMV_i}{\sum_{i=1}^{9} CMV} \tag{2}$$

$$DSDR_{a,i} = \frac{APV_i}{\sum_{i=1}^{9} APV} \tag{3}$$

where,

$DSDR_{c,i}$ = the current discipline-specific distance ratio that reflects on the differences of the CMVs
$DSDR_{a,i}$ = the actual discipline-specific distance ratio that reflects on the differences of the APVs

For example, in the discipline of geology, the average of the top 25 percent individual publication performance value is 49 (48.769) publications in a 5-year period, whereas in the discipline of geodesy/geoinformatics, that value is 31 (30.900) publications. At the moment, the Section defines 30 publications as CMV for both disciplines; however, in reality, the geologists produce 58 percent more publications than the surveyors in the same period. As for the DSDRs, in the case of both disciplines, the current DSDR is 0.107143 (30/280) (i.e., the ratio of the discipline-specific CMV [30] and the total amount of the CMVs [280]), but, taking the APVs into account, the actual DSDR for geologists should be 0.127070 (48.769/383.797), and for surveyors, it should be 0.080511 (30.900/383.797) (see the data in Fig. 1 and Table 3). Based on the method described above, in both cases, the total value of the DSDR is equal to 1. Naturally, the ratios of the RMVs are in line with the actual DSDRs.

Previously, by considering the APVs, we determined the number of years ($Y_i$) required to fulfill the CMVs for each discipline. Based on the results, in the case of each indicator, we can now calculate the mean of $Y_i$.

$$Y_m = \frac{\sum Y_i}{9} \quad (4)$$

By having this information (i.e., the $Y_m$), it becomes possible to optimize the number of years necessary to fulfill the minimum values. Thus, the ratio of $Y_m$ and $Y_i$ must be calculated.

$$R_{y,i} = \frac{Y_m}{Y_i} \quad (5)$$

Now, in the case of each performance indicator and for each discipline, the recalibrated minimum values (RMVs) can be defined.

$$RMV_i = CMV_i * R_{y,i} \quad (6)$$

As a fundamental principle, the ratio of the discipline-specific CMVs per indicator reflects the current DSDR, but the ratio of the discipline-specific RMVs must reflect the actual DSDR.

In conclusion, the RMV can be considered to be a combination of such factors as the CMVs of the performance indicators by disciplines, the number of years required to fulfill the CMVs, and the actual DSDRs derived from the APVs by disciplines. Hence, for geologists, the RMV will be 36 (36.197) publications, and for surveyors, it will be 23 (22.934). Both of them require 3.711 years ($Y_m$) to fulfill the discipline-specific RMVs.

Because publication trends are changing continuously, the APVs must be reviewed regularly (e.g., every five or ten years), and the actual DSDRs must be recalculated as well.

The discipline-specific publication output is influenced by several factors. However, the number of co-authors is considered to be one of the most crucial factors. It is observed that the number of co-authors per publication has been gradually increasing for a long time, but in the case of both natural (particularly in physics) and medical sciences, the magnitude of the increase is much higher than in the case of social sciences (Henriksen, 2016, 2018; Ossenblok, Verleysen, & Engels, 2014). In social sciences, the single-author publications are still particularly common, whereas in physics, a paper with more than 5,000 co-authors has been produced (Castelvecchi, 2015). Therefore, it is of high importance to allocate authorship credit by, for example, employing a fractional counting approach (see, e.g., Bouyssou & Marchant, 2016; Cronin, 2001; De Moya-Anegon, Guerrero-Bote, Lopez-Illescas, & Moed, 2018; Gauffriau, Larsen, Maye, Roulin-Perriard, A., & Von Ins, 2007; Hagen, 2010; Osório, 2018; Van Hooydonk, 1997; Zhou & Leydesdorff, 2011). When using fractional counting, one credit is equally or proportionally shared among co-authors (Lin, Huang, & Chen, 2013; Sivertsen, Rousseau, & Zhang, 2019; Waltman & van Eck, 2015).

As can be seen in Table 2, the distribution of multi-authored publications among different disciplines shows an inhomogeneous pattern: in the case of the discipline of social geography, only two-thirds of the publications are multi-authored, and the average number of authors in multi-authored publications remains under four, whereas 95 percent of the publications in geophysics are created in co-production, and the average number of authors in those publications exceeds 9.5.

**Table 2.** Co-authorship characteristics by disciplines in the Section of Earth Sciences

|  | Number of scientific publications, 2014–2018 | Number of multi-authored scientific publications, 2014–2018 | Ratio of number of scientific publications to number of multi-authored scientific publications (%) | Number of co-authors in multi-authored publications | Average number of co-authors per multi-authored publication |
|---|---|---|---|---|---|
| Geochemistry, Mineralogy and Petrology | 1,608 | 1,529 | 95.09 | 11,166 | 7.30 |
| Geodesy and Geoinformatics | 591 | 444 | 75.13 | 2,135 | 4.81 |
| Geology | 1,154 | 1,073 | 92.98 | 6,533 | 6.09 |
| Geophysics | 889 | 841 | 94.60 | 8,036 | 9.56 |
| Meteorology | 1,064 | 969 | 91.07 | 5,481 | 5.66 |
| Mining | 679 | 588 | 86.60 | 2,724 | 4.63 |
| Paleontology | 532 | 443 | 83.27 | 2,543 | 5.74 |
| Physical Geography | 2,166 | 1,875 | 86.57 | 9,633 | 5.14 |
| Social Geography | 3,277 | 2,199 | 67.10 | 8,173 | 3.72 |

Due to the fact that the co-authorship characteristics significantly differ by disciplines, it is highly important to determine the RMVs by employing both integer and fractional counting methods. That is, regarding each performance indicator being involved in this study, two types of RMVs will be demonstrated.

## 3. Results

### 3.1. Number of scientific publications

In the case of the performance indicator "The number of scientific publications", for each discipline, the Section requires the fulfillment of 30 publications as a minimum value except for the discipline of social geography, for which the minimum value required to be achieved is 40 publications. By examining the top 25 percent of individual publication performance per discipline by employing the integer counting method, we can conclude that the APVs exceed the CMVs in the case of each discipline (Table 3). The CMVs of the performance indicator as being required by the Section are significantly over-fulfilled by the miners, geologists, and meteorologists, and slightly over-fulfilled by the social geographers and surveyors. That is, it takes different amounts of time to achieve a CMV for researchers affiliated with different disciplines. For example, geologists and miners require approximately three years to achieve the CMV, whereas surveyors need five years to do the same. Regarding these performance indicators, the mean of the years ($Y_m$) necessary to fulfill the CMV is 3.711. Considering the value of the $Y_m$ and the actual DSDRs (Fig. 1), the minimum values of the performance indicator can be recalibrated. This means that, for example, the minimum value for geologists increases from 30 (CMV) to 36 publications (RMV), and for surveyors, it decreases from 30 (CMV) to 23 publications (RMV).

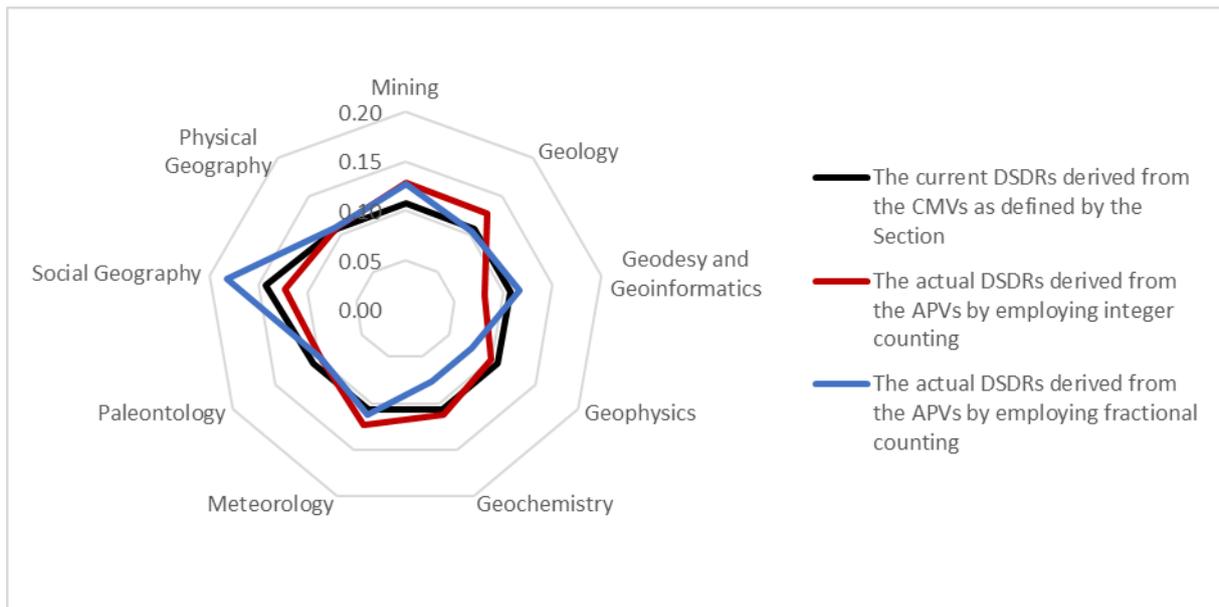

**Fig. 1.** Current DSDRs regarding the performance indicator "Number of scientific publications" and actual DSDRs by employing integer and fractional counting

**Table 3.** CMVs and RMVs regarding the performance indicator "Number of scientific publications"

|  | Current minimum values | Number of scientific publications in HSB, 2014–2018 | | Number of years required to fulfill minimum value | | Recalibrated minimum values | |
| --- | --- | --- | --- | --- | --- | --- | --- |
|  |  | IC* | FC** | IC* | FC** | IC* | FC** |
| Mining | 30 | 49.125 | 18.159 | 3.053 | 8.260 | 36 (36.461) | 13 (13.478) |
| Geology | 30 | 48.769 | 14.774 | 3.076 | 10.153 | 36 (36.197) | 11 (10.965) |
| Geodesy and Geoinformatics | 30 | 30.900 | 16.747 | 4.854 | 8.957 | 23 (22.934) | 12 (12.430) |
| Geophysics | 30 | 38.333 | 10.974 | 3.913 | 13.669 | 28 (28.451) | 8 (8.145) |
| Geochemistry, Mineralogy and Petrology | 30 | 43.200 | 10.999 | 3.472 | 13.637 | 32 (32.063) | 8 (8.164) |
| Meteorology | 30 | 47.538 | 16.176 | 3.155 | 9.273 | 35 (35.283) | 12 (12.006) |
| Paleontology | 30 | 37.375 | 14.030 | 4.013 | 10.691 | 28 (27.740) | 10 (10.413) |
| Social Geography | 40 | 46.972 | 26.053 | 4.258 | 7.677 | 35 (34.863) | 19 (19.337) |
| Physical Geography | 30 | 41.583 | 15.610 | 3.607 | 9.610 | 31 (30.863) | 12 (11.585) |

*IC: calculated by integer counting method; **FC: calculated by fractional counting method

If employing the fractional counting method, based on the top 25 percent of individual publication performance, social geographers require an average of 7.677 years to achieve the CMV, which is the lowest average in the Section (Table 3). The reason for this result is that the discipline of social geography is characterized by the lowest ratio of multi-authored publications and the lowest average number of authors in multi-authored papers. In contrast, in the case of the disciplines of geophysics and geochemistry, which are both characterized by the highest ratio of multi-authored publications and the highest average number of authors in those multi-authored papers, approximately 14 years is necessary to achieve the CMV. However, as a fundamental principle, the time interval to achieve the RMVs must be adjusted to be exactly the same if the APVs are calculated by both integer and fractional counting methods. That is, regarding the performance indicator "The number of scientific publications", the $Y_m$ must be 3.711 years, for each discipline equally. Considering the actual DSDRs derived from the APVs by employing the fractional counting method and the $Y_m$ value, for social geographers, the RMV becomes the highest in the section with 19 (19.337) publications, and for the disciplines of geophysics and geochemistry, it reduces to 8 (8.145 and 8.164) publications. In the case of the disciplines of geodesy/geoinformatics and social geography, the ratio of RMVs calculated by both integer and fractional counting will remain under two, whereas in geochemistry and geophysics, it will approximate four.

**3.2. Number of journal articles indexed in Web of Science**

The international visibility of researchers affiliated with particular disciplines in terms of the number of journal articles indexed in international databases (in this case, in WoS) significantly differs. Table 4 demonstrates that, based on the top 25 percent of individual publication performance, the average number of WoS-indexed articles in the disciplines of geodesy/geoinformatics, mining, and social geography is relatively low, at least, if comparing with data of other disciplines. For example, the number of WoS-indexed articles produced in a 5-year period by the top social geographers is approximately one-fourth what geochemists produce in the same period. Due to multiple reasons, social geographers (surveyors and miners as well) do not publish articles in WoS-indexed journals as frequently as researchers affiliated with other earth science disciplines. It has been demonstrated in many studies that journal articles written in English and journal articles produced in the fields of natural and medical sciences are significantly overrepresented in WoS (in this respect, Scopus is similar to WoS). In social sciences, however, the books and book chapters are important channels of scientific communication as well, and in the case of social geography (and physical geography), the maps are also considered to be scientific publications. In addition, in the disciplines of geodesy/geoinformatics (as is also the situation for computer sciences), conferences are also important dissemination routes of knowledge (see, e.g., Vrettas & Sanderson, 2015). For the hard natural sciences (e.g., physics and chemistry), journal articles are considered to be the standard way to communicate new information. Furthermore, in the disciplines of geochemistry, geology, geophysics, and meteorology, research projects are quite often carried out by international research teams with many participants, and those large-scale collaborations require the research outcome to be published in international journals.

     By employing the integer counting method, the average number of WoS-indexed articles produced by the top 25 percent of geochemists and geologists in a 5-year period reaches the highest value in the Section. In the case of the discipline of geochemistry, the APV approximates 20 articles (i.e., four articles per year). In fact, the CMVs more or less reflect the differences: The Section requires researchers affiliated with the disciplines of geochemistry, geology, geophysics, meteorology, and paleontology to fulfill a much higher minimum value (12 articles) than researchers in the disciplines of mining, geodesy/geoinformatics, and physical geography (eight articles). To achieve the CMV, social geographers must produce only six WoS/Scopus-indexed articles. In addition, taking the APVs into account, the first group seems to be rather inhomogeneous because, for example, geochemists produce 62 percent more WoS indexed-articles than geophysicists. Consequently, the top geochemists fulfill the CMV in approximately three years, whereas for the top geophysicists, it takes five years. Surveyors and social geographers, however, are only able to achieve the CMV in six years (more precisely: 5.882 and 5.775 years), despite the fact that for them, the CMV is 33 percent and 50 percent lower than for geochemists. In the discipline of mining, for top researchers, an average of 7.4 years is required to fulfill the CMV; that is, twice as many as geologists need.

     When recalibrating the minimum values of the performance indicator "The number of journal articles indexed in the Web of Science", for each discipline, the number of years required to fulfill the RMVs ($Y_m$) must be 4.715. Considering the actual DSDRs (Fig. 2), the CMV for miners and social geographers should be reduced to five articles, respectively, and for geochemists and geophysicists, rather than being 12–12, the CMV should be increased to 19 and 17 (Table 4).

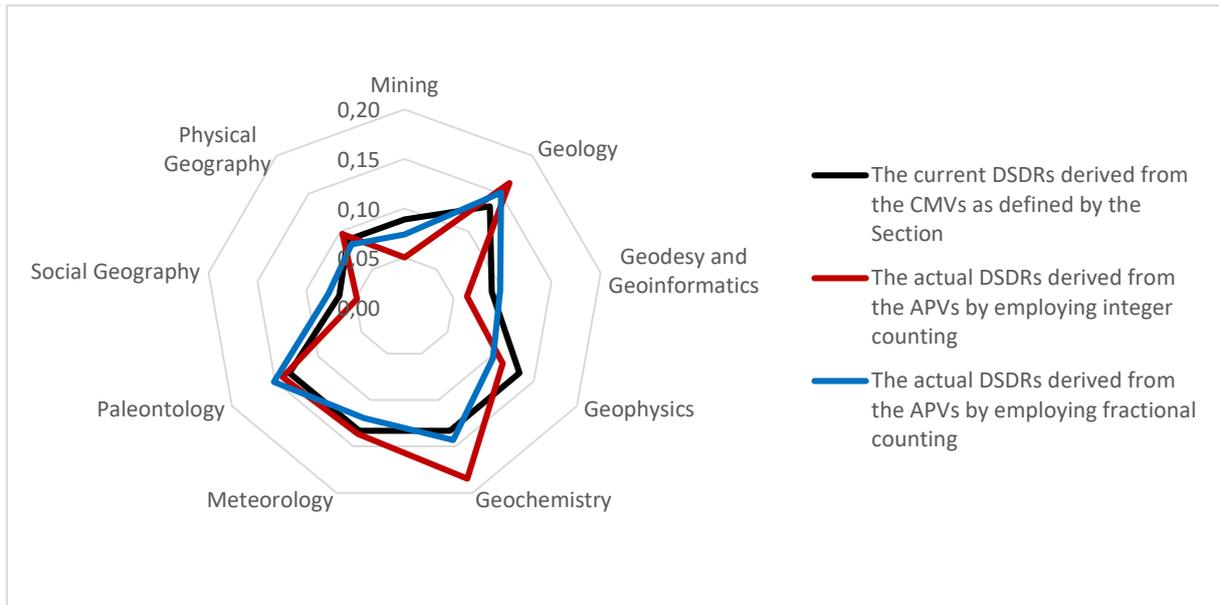

**Fig. 2.** Current DSDRs regarding the performance indicator "Number of journal articles indexed in the Web of Science" and actual DSDRs by employing integer and fractional counting

**Table 4.** CMVs and RMVs regarding the performance indicator "Number of journal articles indexed in the Web of Science"

|  | Current minimum values | Number of WoS-indexed journal articles, 2014–2018 | | Number of years required to fulfill minimum value | | Recalibrated minimum values | |
|---|---|---|---|---|---|---|---|
|  |  | IC* | FC** | IC* | FC** | IC* | FC** |
| Mining | 8 | 5.375 | 1.968 | 7.442 | 20.326 | 5 (5.069) | 2 (1.856) |
| Geology | 12 | 17.538 | 4.046 | 3.421 | 14.830 | 17 (16.539) | 4 (3.815) |
| Geodesy and Geoinformatics | 8 | 6.800 | 2.611 | 5.882 | 15.318 | 6 (6.412) | 2 (2.462) |
| Geophysics | 12 | 12.167 | 2.751 | 4.932 | 21.810 | 11 (11.473) | 3 (2.594) |
| Geochemistry, Mineralogy and Petrology | 12 | 19.750 | 3.833 | 3.038 | 15.652 | 19 (18.624) | 4 (3.615) |
| Meteorology | 12 | 14.615 | 3.196 | 4.105 | 18.774 | 14 (13.782) | 3 (3.014) |
| Paleontology | 12 | 15.125 | 4.054 | 3.967 | 14.801 | 14 (14.263) | 4 (3.823) |
| Social Geography | 6 | 5.194 | 2.080 | 5.775 | 14.420 | 5 (4.898) | 2 (1.962) |
| Physical Geography | 8 | 10.333 | 2.221 | 3.871 | 18.011 | 10 (9.744) | 2 (2.094) |

**IC: calculated by integer counting method; **FC: calculated by fractional counting method

If employing the fractional counting method, the RMVs will reduce as compared to the RMVs calculated by integer counting. As has been demonstrated earlier, it is the social geographers who produce the lowest ratio of multi-authored publications and involve the lowest average number of co-authors in those publications. Hence, in the case of the social geography discipline, the fractionally counted RMV will reduce by 150 percent, whereas the disciplines of geochemistry and physical geography experience 375 percent and 400 percent reductions, respectively.

### 3.3. Number of independent citations

The independent citations are those that neither include self-citations nor citations received from co-authors. To recalibrate the minimum values of this indicator, the independent citations (henceforward: citations) received in the period of 2014−2019 by publications produced in the period of 2014−2018 has been taken into account (this means that $t = 6$). The Section defines 120 citations as the CMV for researchers affiliated with the disciplines of mining, geodesy/geoinformatics, and physical geography, and 150 citations for all other disciplines. As can be seen in Table 5, in a 6-year period, the highest number of citations were received by meteorologists and paleontologists (of course, taking the top 25 percent of individual publication performance into account). For them, it took less than five years to achieve the CMV, whereas surveyors and geophysicists required more than 13 and 15 years, respectively. By employing the integer counting method, the mean of the years necessary to fulfill the performance indicator "The number of independent citations" will be 8.110. If considering the combination of the CMVs, the $Y_m$ and the actual DSDR (Fig. 3), the CMVs for the researchers affiliated with the disciplines of geodesy/geoinformatics and geophysics will significantly decrease (for geophysicists, by about 50 percent). Other disciplines will experience increases regarding the CMV; for example, for social geographers, the CMV will increase by 10 percent, whereas for meteorologists and paleontologists, it will increase by approximately 60 percent.

When employing the fractional counting method, it turns out that for the disciplines that regularly produce a high ratio of multi-authored publications with many co-authors in those publications, the number of years required to fulfill the CMV would be unrealistically high (e.g., for geophysicists, it would take 76 years to fulfill the CMV). It is essential, however, that the time interval to achieve the minimum value be the same (i.e., 8.110 years) for each discipline, irrespective of whether fractional or integer counting methods are used. Therefore, the RMV for the disciplines of geophysics and geodesy/geoinformatics will reduce to 16 citations, respectively. The social geographers, however, will experience a less significant decrease in the CMV calculated by the integer counting method: from 165 citations, the RMV will reduce to 70 citations (which is 42 percent of the former). As can be seen in Fig. 3., if using fractional counting, it will be the social geographers who have to produce the highest number of citations as RMVs in the Section.

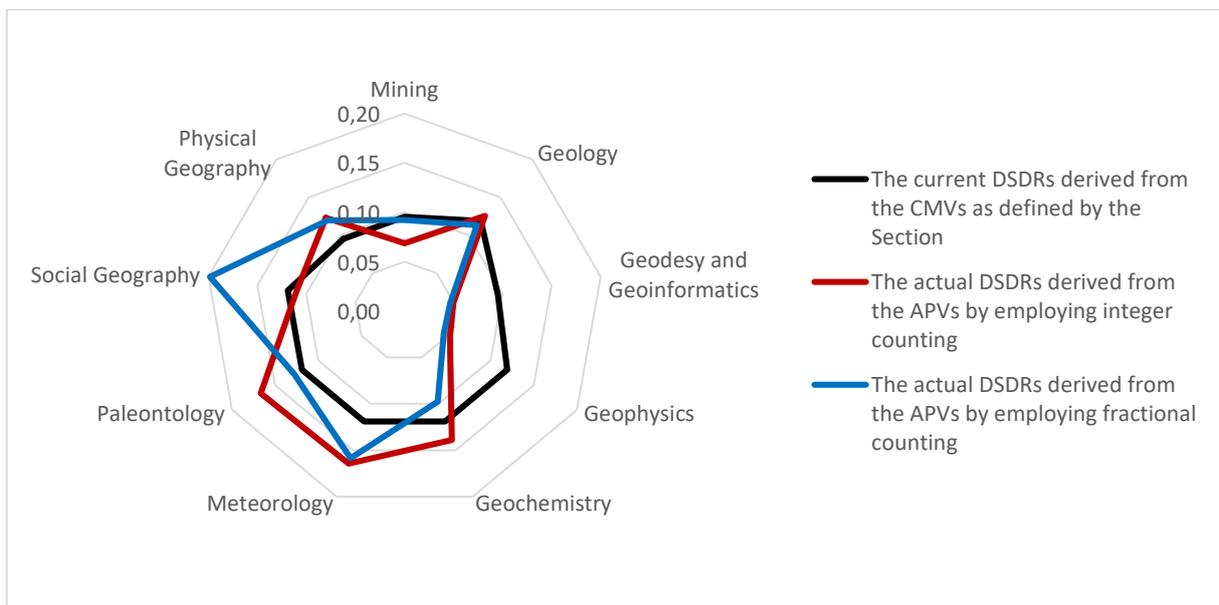

**Fig. 3.** Current DSDRs regarding the performance indicator "Number of independent citations" and actual DSDRs by employing integer and fractional counting

**Table 5.** CMVs and RMVs regarding the performance indicator "Number of independent citations"

|  | Current minimum values | Number of independent citations, 2014–2019 | | Number of years required to fulfill minimum value | | Recalibrated minimum values | |
|---|---|---|---|---|---|---|---|
|  |  | IC* | FC** | IC* | FC** | IC* | FC** |
| Mining | 120 | 62.875 | 19.934 | 9.543 | 30.099 | 102 (101.983) | 32 (32.333) |
| Geology | 150 | 115.538 | 24.531 | 6.491 | 30.573 | 187 (187.403) | 40 (39.790) |
| Geodesy and Geoinformatics | 120 | 45.800 | 10.132 | 13.100 | 59.221 | 74 (74.288) | 16 (16.433) |
| Geophysics | 150 | 48.583 | 9.845 | 15.437 | 76.179 | 79 (78.802) | 16 (15.969) |
| Geochemistry, Mineralogy and Petrology | 150 | 127.700 | 21.111 | 5.873 | 35.526 | 207 (207.129) | 34 (34.242) |
| Meteorology | 150 | 151.154 | 34.312 | 4.962 | 21.858 | 245 (245.172) | 56 (55.654) |
| Paleontology | 150 | 153.250 | 27.642 | 4.894 | 27.133 | 249 (248.572) | 45 (44.835) |
| Social Geography | 150 | 101.583 | 42.932 | 7.383 | 17.470 | 165 (164.768) | 70 (69.636) |
| Physical Geography | 120 | 113.167 | 25.925 | 5.302 | 23.144 | 184 (183.556) | 42 (42.050) |

*IC: calculated by integer counting method; **FC: calculated by fractional counting method

In addition, except for the discipline of social geography, the performance indicator "Number of independent citations located in the WoS (more precisely: in the SCI/SSCI) and the Scopus" is included in the performance indicator list of each other discipline belonging to the Section (see, Table 1). The current structure of the performance indicators (i.e., their types and minimum values) has been effective since 2012, when the Section modified the previous requirements of the DSc title. Until 2012, social geographers were required to obtain at least 15 independent citations located in SCI/SSCI/Scopus. When the scientific requirements of the application for the DSc title were reframed by senior researchers of the Committee on Social Geography, they removed that indicator type from the collection of the performance indicators. If considering the fact that only 17 percent of the citations being received by the publications of social geographers comes from WoS-indexed journals, the above action seems to be quite reasonable. In contrast, for example, in the case of the disciplines of geochemistry and geology, the ratio of the total number of citations to the citations in WoS-indexed journals is approximately 70 percent. The difference between the disciplines regarding the value of this particular indicator seems to be too large, and such a high difference would result in bias when evaluating applications for the DSc title. The discipline of mining is also characterized by a low ratio of citations in WoS-indexed journals; however, because mining has been placed into the same group with the disciplines of geodesy/geoinformatics and physical geography, the Committee on Mining cannot ignore the use of this type of performance indicator.

### 3.4. Cumulative impact factor value

In spite of the fact that the journal impact factor (JIF) was originally created to be a tool that helps evaluate journals (Garfield, 1972), since the date of its creation, it has gained widespread popularity as an indicator demonstrating individuals' research performance. According to Garfield (2006: 92), however, "the use of journal impacts in evaluating individuals has its inherent dangers." In fact, there are ongoing debates among experts in bibliometrics regarding whether the JIF is suitable for demonstrating individuals' research performance (see, for example, Alberts, 2013; Buela-Casal & Zych, 2012; Seglen, 1997; Waltman & Traag, 2017). Recently, some major efforts have emerged to eliminate the use of the JIF in funding, appointment, and promotion considerations (the most well-known is the San Francisco Declaration on Research Assessment [DORA]) (Cagan, 2013; Zhang, Rousseau, & Sivertsen, 2017), and reframe the entire research evaluation process by giving less significance to metric-based approaches (see the Leiden Manifesto) (Hicks, Wouters, Waltman, De Rijcke, & Rafols, 2015). Irrespective of the fact that the efforts to eliminate the JIF are rapidly spreading in the international scientific community, in Hungary, the JIF is still considered to be the cornerstone in evaluating individuals' research performance. The JIF is one of the most important performance indicators during the evaluation of applications for scientific qualifications (i.e., DSc title, Habilitation, and PhD degree), promotions at universities and research institutes, and funding applications.

Naturally, the JIF is a fundamental indicator for the Section of Earth Sciences as well (there are rumors, however, that the Section is considering the removal of the JIF from among the performance indicators). In the case of the hard natural sciences (e.g., geochemistry, geology, and geophysics), the CMV, regarding the performance indicator "The cumulative impact factor value" is eight; that is, researchers are required to publish articles in such journals that have a cumulative impact factor of at least eight. In the case of the disciplines of mining, geodesy/geoinformatics, and physical geography, researchers have to produce a cumulative impact factor value of four, and for social geographers, the Section indicates a cumulative impact factor value of two only as CMV.

From an international perspective, it may seem surprising that for social geographers, such a low cumulative impact factor value has been defined, but the truth is that in the period of 2014−2018, 64 percent of the social geographers did not publish a single article in journals being listed in the Journal Citation Reports (JCR). In contrast, in the discipline of geochemistry, only 9 percent of the researchers had at least one article published in JCR-listed journals. In addition, the low cumulative impact factor value having been introduced for social geographers is in line with the general observation that suggests that the average JIF values vary across fields, but the lowest average JIF value is the characteristic of social sciences (Nature Index, 2018). We can also conclude that by introducing such a low cumulative impact factor value for social geographers, the Section indirectly reinforced the fact that the discipline of social geography rather belongs to the field of social sciences than to natural sciences.

Table 6 demonstrates that in the case of each discipline, the APVs exceed the CMVs. The disciplines of social geography and mining produce approximately 100 percent more cumulative impact factor value than is required; in the disciplines of physical geography, meteorology, and paleontology, the average over-fulfillment ranges from 360 to 400 percent, whereas in the case of the disciplines of geology and geochemistry, the CMV is exceeded by 445 and 550 percent, respectively. As has been demonstrated, the differences between the discipline-specific APVs regarding the performance indicator "The cumulative impact factor value" are too high; therefore, it is not surprising that some scholars urge the Section to eliminate the impact factor in evaluating individuals' research performance. As a matter of fact, by introducing the discipline-specific CMVs, the Section attempted to equilibrate (lessen at least) the differences. For example, geochemists produce 13 times higher APV than social geographers (see the actual DSDRs in Fig. 4), but due to the fact that geochemists are required to fulfill a higher minimum value, the social geographers need only three times as many years to achieve the CMV than do geochemists.

If we employ the integer counting method to recalibrate the minimum values, the mean of the years needed to fulfill the performance indicator "The cumulative impact factor value" will be 1.430. Considering this mean-year value, for social geographers, the CMV will reduce from two to one (1.174), and for miners from four to three (2.521), for geophysicists it will remain eight (8.049), and for geochemists, the CMV will increase from eight to 15 (14.840).

If we define the RMVs by using fractional counting, the cumulative impact factor values will be much lower for each discipline (Table 6). For example, in this respect, the cumulative impact factor value for geochemists will be only 7.4 times higher than for social geographers. As can be seen in Table 6, the social geographers would experience the lowest RMV with a cumulative impact factor value of 0.343, whereas, it is the geologists who would be required to produce the highest cumulative impact factor value (2.608).

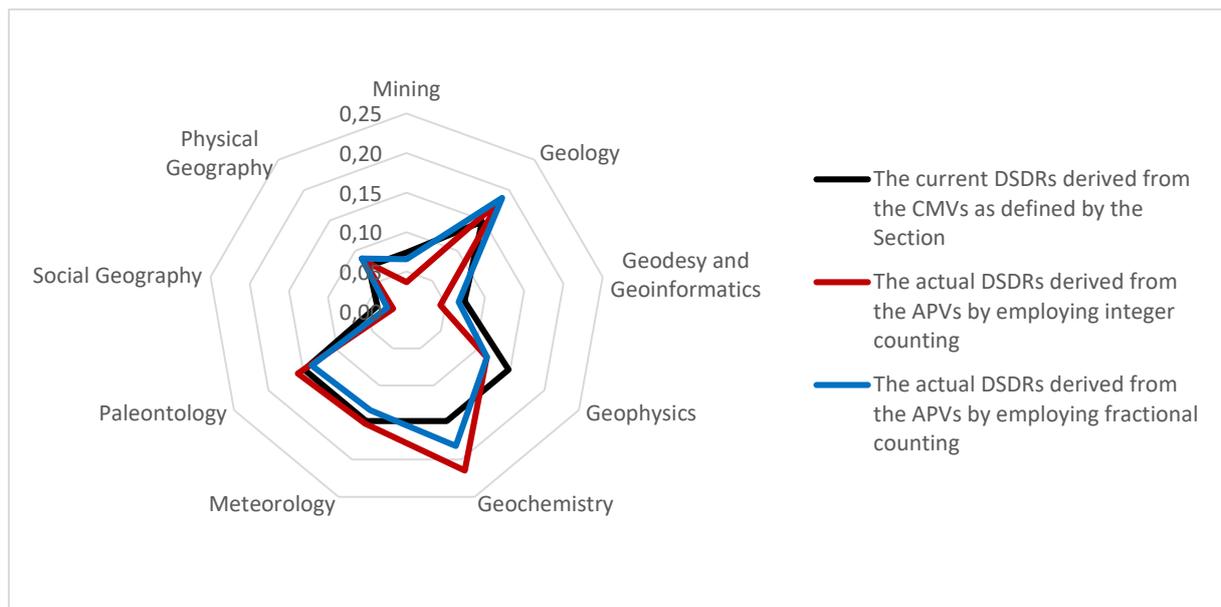

**Fig. 4.** Current DSDRs regarding the performance indicator "The cumulative impact factor value" and actual DSDRs by employing integer and fractional counting

**Table 6.** CMVs and RMVs regarding the performance indicator "The cumulative impact factor value"

|  | Current minimum values | Cumulative impact factor, 2014–2018 | | Number of years required to fulfill minimum criteria | | Recalibrated minimum values | |
|---|---|---|---|---|---|---|---|
|  |  | IC* | FC** | IC* | FC** | IC* | FC** |
| Mining | 4 | 8.815 | 3.202 | 2.269 | 6.247 | 3 (2.521) | 0.916 |
| Geology | 8 | 43.677 | 9.120 | 0.916 | 4.386 | 12 (12.492) | 2.608 |
| Geodesy and Geoinformatics | 4 | 10.484 | 3.253 | 1.908 | 6.147 | 3 (2.998) | 0.930 |
| Geophysics | 8 | 28.144 | 5.695 | 1.421 | 7.023 | 8 (8.049) | 1.629 |
| Geochemistry, Mineralogy and Petrology | 8 | 51.888 | 8.854 | 0.771 | 4.518 | 15 (14.840) | 2.532 |
| Meteorology | 8 | 36.650 | 6.498 | 1.091 | 6.156 | 10 (10.482) | 1.858 |
| Paleontology | 8 | 38.067 | 6.710 | 1.051 | 5.961 | 11 (10.887) | 1.919 |
| Social Geography | 2 | 4.104 | 1.198 | 2.436 | 8.346 | 1 (1.174) | 0.343 |
| Physical Geography | 4 | 19.950 | 4.246 | 1.002 | 4.710 | 6 (5.706) | 1.214 |

*IC: calculated by integer counting method; **FC: calculated by fractional counting method

## 4. Discussion and Conclusion

Due to multiple reasons, reforming the academic performance indicators of the DSc title, the highest scientific qualification in Hungary being awarded by the Hungarian Academy of Sciences, seems unavoidable. The demand for introducing reforms is in line with some recently emerged trends in international science (e.g., the growing popularity of open-access publishing and the efforts being initiated to eliminate the use of the journal impact factor). In general, the academic performance indicators have two crucial factors that may need to be changed: their types and their minimum values.

In this paper, a new method was presented regarding how the minimum values of the performance indicators were most optimally recalibrated and as a case study, the Section of Earth Sciences of the HAS was chosen. To achieve the research goal, a straightforward bibliometric analysis was conducted, revealing the individual publication performance of each researcher affiliated with the Section. The Section of Earth Sciences encompasses nine scientific committees, each of which represents a particular scientific discipline. Due to the fact that most disciplines of the Section are considered to be hard natural sciences, whereas social geography rather belongs to the field of social sciences, the scientific profile of the section can be characterized by high inhomogeneity. However, if considering the current types and values of the performance indicators, we can conclude that this inhomogeneity is not adequately acknowledged by the Section, and it may generate tension among researchers during the evaluation of their research performance. The results of this analysis allowed us to recalibrate the minimum values of the performance indicators by taking the discipline-specific differences into account.

First, after scrutinizing the actual performance values of each researcher, it turns out that taking a particular time period into account, the publication output of researchers having a DSc title rather approximates the average. For this reason, during the recalibration of the minimum values of the performance indicators, a new reference group, the top 25 percent of individual publication performance per indicator and discipline must be chosen. After obtaining information on the actual performance values (derived from the top 25 percent of individual publication performance), we can conclude that the actual discipline-specific distance ratios significantly differ from the current discipline-specific distance ratios (which come from the differences between the current minimum values being defined by the Section). Furthermore, it also turns out that in the case of some disciplines (primarily those that are considered to be hard natural sciences), the ratio of multi-authored publications, and the number of co-authors in those publications are extremely high, but the Section neglects to attach importance to those facts. Thus, when a researcher applies for a DSc title (or a university professorship position), he/she in fact demonstrates the publication output of research teams instead of demonstrating his/her own individual publication performance (e.g., a publication with 100 co-authors counts as one credit for each author). Due to the fact that the Section disregards the use of the fractional counting method during the evaluation of individuals' publication performance, a strong bias is experienced towards researchers affiliated with hard natural science disciplines.

To recalibrate the minimum value of the performance indicators, two factors must be considered: 1) the actual discipline-specific distance ratios, and 2) the number of co-authors in multi-authored publications per discipline. Taking the first factor into account, the recalibrated minimum values must reflect the actual discipline-specific distance ratio being based on the actual performance values derived from the top 25 percent of individual publication performance. By doing this, we can harmonize the time interval required to achieve the minimum value of the performance indicators for each researcher, irrespective of which discipline he/she is affiliated with. Considering the second factor, it is recommended that the Section employs the fractional counting method when evaluating an individual's publication performance. Naturally, the fractional counting approach has its critics as well because it does not attribute importance to the authorship order (e.g., it erodes the significance of the first and last author position) (see, for example, Egghe et al., 2000; Todeschini & Baccini, 2016; Tscharntke, Hochberg, Rand, Resh, & Krauss, 2007; Vavryčuk, 2018). However, by using fractional counting, the authorship credit can be equally shared among co-authors, allowing evaluators not to consider teamwork as individual contribution. In conclusion, the minimum values of the performance indicators of the DSc title would be more harmonically recalibrated if the above recommendations were considered (Appendix 1 demonstrates the recalibrated minimum values for each discipline by employing integer counting method).

In addition, the types of performance indicators should be more carefully redefined. In tandem with the effort of the international scientific community to eliminate the use of the journal impact factor in the evaluation of individuals' research performance, the Section should consider removing "The cumulative impact factor value" from among the performance indicators. For social geographers, however, the performance indicator "The number of independent citations located in the WoS and the Scopus", which is currently not part of the requirements, should be re-included. This suggestion is based on the following hypotheses: It is either the case that researchers affiliated with certain disciplines produce low-performance indicator values because this is a special feature of their discipline, or they are simply not required to produce higher values. More precisely: Why is it that social geographers produce a low number of articles in WoS-indexed journals as compared to that of researchers affiliated with other disciplines? Is it that this is a special feature of the discipline of social geography, or is it because social geographers are not required (i.e., motivated) to produce a higher number of WoS-indexed articles (even if they could do it)? Naturally, this hypothesis should be tested carefully before changes are introduced.

Finally, it would be important to avoid merging disciplines into groups without thoroughly analyzing the publication and citation characteristics of researchers affiliated with those disciplines. This paper demonstrates that researchers belonging to particular disciplines might have highly different publication performances from those they are now being grouped with. Thus, when applying for a research grant, a researcher affiliated with a discipline (e.g., geology) has to compete not only with his/her professional peers, but also researchers affiliated with other disciplines (e.g., geochemistry), who might have more or less different publication and citation characteristics.

In conclusion, in the Section of Earth Sciences, neither the types nor the minimum values of the current performance indicators of the DSc title are correctly defined, and this issue may bias the outcome of the evaluation in individuals' research performance.

**Appendix 1**

Recalibrated minimum values for disciplines belonging to the Section of Earth Sciences*

| | Geochemistry, Mineralogy, and Petrology | Geodesy and Geoinformatics | Geography I (Social Geography) | Geography II (Physical Geography) | Geology | Geophysics | Meteorology | Mining | Paleontology |
|---|---|---|---|---|---|---|---|---|---|
| Number of scientific publications | 32 (+2) | 23 (-7) | 35 (-5) | 31 (+1) | 36 (+6) | 28 (-2) | 35 (+5) | 36 (+6) | 28 (-2) |
| Number of scientific publications with first author position | 16 (+1) | 12 (-3) | 18 (-2) | 16 (+1) | 18 (+3) | 14 (-1) | 18 (+3) | 18 (+3) | 14 (-1) |
| Number of scientific publications since obtaining last scientific degree | 16 (+1) | 12 (-3) | 26 (-4) | 16 (+1) | 18 (+3) | 14 (-1) | 18 (+3) | 18 (+3) | 14 (-1) |
| Number of scientific books and monographies | - | - | 2 (-) | - | - | - | - | - | - |
| Number of scientific publications published in a foreign language | - | - | 31 (-4) | - | - | - | - | - | - |
| Number of journal articles indexed in SCI/SSCI and Scopus | 19 (+7) | 6 (-2) | 5 (-1) | 10 (+2) | 17 (+5) | 11 (-1) | 14 (+2) | 5 (-3) | 14 (+2) |
| Number of journal articles indexed in SCI/SSCI and Scopus since obtaining last scientific degree | 10 (+4) | 3 (-1) | 3 (-) | 5 (+1) | 9 (+3) | 6 (-) | 7 (+1) | 3 (-1) | 7 (+1) |
| Number of independent citations | 207 (+57) | 74 (-46) | 165 (+15) | 184 (+64) | 187 (+37) | 79 (-71) | 245 (+95) | 102 (-18) | 249 (+99) |
| Number of independent citations located in SCI/SSCI and Scopus | 82 (+32) | 24 (-6) | 18 (newly introduced) | 53 (+23) | 74 (+24) | 27 (-23) | 74 (+24) | 15 (-15) | 94 (+44) |
| Cumulative impact factor value | 15 (+7) | 3 (-1) | 1 (-1) | 6 (+2) | 12 (+4) | 8 (-) | 10 (+2) | 3 (-1) | 11 (+3) |

* In the brackets, differences to the current minimum values are provided.